\documentclass[a4paper, 11pt]{article}
\usepackage[english]{babel}
\makeatletter

\usepackage{geometry,graphicx,color}
\usepackage{amsfonts, amsmath, amssymb, slashed,dsfont}
\usepackage{tensor}
\usepackage{hyperref}
\usepackage{epsfig}
\usepackage{pifont}
  \usepackage{soul}
\usepackage{enumitem}
\usepackage{csquotes}
\usepackage{bm}
\usepackage{mathrsfs} % Allows \mathscr
\usepackage[hyperref,thmmarks,amsmath]{ntheorem}
\usepackage{amsmath,amssymb,slashed}
\usepackage{comment}
\usepackage{tikz}
\usetikzlibrary{positioning}
\usetikzlibrary{decorations.markings}
\usetikzlibrary{arrows.meta}
\usepackage{filecontents}
\usepackage{comment}
\numberwithin{equation}{section}
\newcommand\bbone{\mathbb{I}}

\newcommand\kS{{\mathfrak S}}

\newcommand{\sign}{\mathrm{sign}}
\newcommand\phib{\overline{\phi}}
\newcommand{\institute}[1]{\newcommand{\@institute}{#1}}
\renewcommand{\maketitle}{
\vspace*{0.5\baselineskip}
{% title
\center\LARGE\noindent\@title\par
}%
\vspace{1.5\baselineskip}
{% author
\center\normalsize\noindent\ignorespaces\@author\par
}%
\vspace{0.5\baselineskip}
{% institutequantum
\center\normalsize\ignorespaces\@institute\par
}%
\vspace{2\baselineskip}
}%

\usepackage{siunitx}
\usepackage{xcolor}
\usepackage{booktabs,colortbl, array}
\usepackage{pgfplotstable}
\usepackage{pbox}
\definecolor{rulecolor}{RGB}{0,71,171}
\definecolor{tableheadcolor}{gray}{0.92}
\newcommand{\topline}{ %
        \arrayrulecolor{rulecolor}\specialrule{0.1em}{\abovetopsep}{0pt}%
        \arrayrulecolor{tableheadcolor}\specialrule{\belowrulesep}{0pt}{0pt}%
        \arrayrulecolor{rulecolor}}
\newcommand{\midtopline}{ %
        \arrayrulecolor{tableheadcolor}\specialrule{\aboverulesep}{0pt}{0pt}%
        \arrayrulecolor{rulecolor}\specialrule{\lightrulewidth}{0pt}{0pt}%
        \arrayrulecolor{white}\specialrule{\belowrulesep}{0pt}{0pt}%
        \arrayrulecolor{rulecolor}}
\newcommand{\bottomline}{ %
        \arrayrulecolor{white}\specialrule{\aboverulesep}{0pt}{0pt}%
        \arrayrulecolor{rulecolor} %
        \specialrule{\heavyrulewidth}{0pt}{\belowbottomsep}}%

\pgfplotstableset{normal/.style ={%
        header=true,
        string type,
        column type=l,
        every odd row/.style={
            before row=
        },
        every head row/.style={
            before row={\topline\rowcolor{tableheadcolor}},
            after row={\midtopline}
        },
        every last row/.style={
            after row=\bottomline
        },
        col sep=&,
        row sep=\\
    }
}

\begin{document}

\title{Gauge Theories on quantum Minkowski spaces: \\$\rho$ versus $\kappa$}
%\title{}
\author{Jean-Christophe Wallet}
\institute{%
\textit{IJCLab, Universit\'e Paris-Saclay, CNRS/IN2P3, 91405 Orsay, France}

e-mail:  
\href{mailto:jean-christophe.wallet@universite-paris-saclay.fr}{\texttt{jean-christophe.wallet@universite-paris-saclay.fr}}
}%

\maketitle

\abstract{The $\rho$-Minkowski space-time, a Lie-algebraic deformation of the usual Minkowski space-time is considered. A star-product realization of this quantum space-time together with the characterization of the deformed Poincar\'e symmetry acting on it are presented. It is shown that appearance of UV/IR mixing is expected already in scalar field theories on $\rho$-Minkowski. Classical and one-loop features of a typical gauge theory on this quantum space-time are presented and critically compared to the situation for $\kappa$-Minkowski.}

%% \tableofcontents

%\maketitle
\vfill\eject

\section{Introduction}

The ultimate goal of Quantum Gravity is to provide a description of gravity at ultra short distance and high energy. This challenging attempt has generated a lot of theoretical as well as phenomenological and experimental efforts, see e.g. \cite{zerevue}, \cite{whitepaper}. Among many theoretical expectations has emerged a rather widely accepted consensus that Quantum Gravity may likely imply a quantum space-time at some effective regime. Quantum space-times, which can be described using noncommutative geometry tools \cite{connes}, have thus received a considerable attention and especially those on which acts a deformation of the Poincar\'e symmetry viewed as the quantum space-time symmetry, the deformation parameter being interpreted as the Planck mass or related to the scale of Quantum Gravity. \\

Quantum space-times with "Lie algebra noncommutativity" have been intensively considered from various viewpoints, including studies of the so-called noncommutative field theories \cite{szabo} and noncommutative gauge theories \cite{gauge-rev}. Among them is the very popular $\kappa$-Minkowski space, a deformation of the Minkowski space-time, introduced a long ago \cite{luk-ruegg}, \cite{majid1}, which has generated a huge literature \cite{luk-rev} in view of its possible physical interest, for the Double Special Relativity \cite{amel-nature}, \cite{kow-glik} or Relative 
Locality \cite{rel-loc1, rel-loc2, rel-loc3}.\\

Another deformation of the Minkowski space-time, called the $\rho$-Minkowski space-time, first introduced almost two decades ago in \cite{Lukierski_2006}, has been considered more recently in the context of black-hole physics \cite{marija1}, \cite{marija2}, localisability and quantum observers \cite{localiz1}, \cite{localiz2} and has even emerged in the ADS/CFT context, see \cite{meier1}, \cite{Meier:2023lku}. For a recent exploration of the related algebraic structures, see \cite{fabiano}. One-loop properties of scalar field theories with quartic interaction based on different star-products were studied in \cite{rho3} and \cite{rho-weyl}. The corresponding coordinates Lie algebra in its 3-dimensional version is
\begin{equation}
  [x_0, x_1] = i \rho x_2,\ [x_0, x_2] = - i \rho x_1,\ [x_1, x_2] = 0\label{rho-3d}  
\end{equation}
where the deformation parameter $\rho$ has the dimension of a length, which can be supplemented by a central coordinate, says $x_3$, to reach a 4-d description. A family of Lie-algebraic deformations of the Minkowski space-time stemming from the old classification of the Poisson structures of the Poincar\'e group  \cite{zakrzew} have been introduced \cite{Lukierski_2006} but poorly explored so far. The general form of the corresponding commutation relations between coordinates is given by \begin{equation}
    [x^\mu,x^\nu]=i\zeta^\mu(\eta^{\mu\beta} x^\alpha-\eta^{\nu\beta} x^\alpha)-i\zeta^\nu(\eta^{\mu\beta} x^\alpha-\eta^{\nu\beta} x^\alpha),
\end{equation}
where $\zeta^\mu$ is a vector with length dimension, $\alpha$ and $\beta$ fixed. It turns out that the $\rho$-Minkowski space-time belongs to this family as the algebra \eqref{rho-3d} is recovered for $\zeta^\mu=\delta^\mu_0 $, $\alpha=1$, $\beta=2$.\\

The purpose of this note is to summarize recent results on the description of $\rho$-Minkowski quantum space-time and on scalar field theories and gauge theory model on this quantum space-time, mainly based on 
\cite{rho-weyl}, \cite{gauge-rho}. Section \ref{section2} deals with the construction of the star-product based on the Weyl quantization map combined with
properties of the convolution algebra of the Lie group linked to the coordinates Lie algebra. In Section \ref{section3}, the deformation of the Poincar\'e algebra
acting on the $\rho$-Minkowski space-time. Section \ref{section4} examines the UV/IR mixing occurrence in scalar field theories. In Section \ref{section5}, the construction of a gauge theory on $\rho$-Minkowski is presented, using a twisted differential calculus based on the derivations of the algebra modeling the quantum spate-time and a twisted version of the by-now standard noncommutative version of the Kozsul connection \cite{koszul} already used in \cite{MW2020a}, \cite{MW2021}. The results are discussed in Section \ref{section6}.

\section{Star-product and involution for $\rho$-Minkowski}\label{section2}
A convenient way to construct a star-product realization of the $\rho$-Minkowski space is to combine the Weyl quantization map $Q$ to the convolution algebra $\mathbb{C}(\mathcal{G}):=(L^1(\mathcal{G}),\circ,^*)$ for the Lie group $\mathcal{G}$ related to the coordinates Lie algebra, assuming that any function  $F\in\mathbb{C}(\mathcal{G})$ is a function on the momentum space: one has $F=\mathcal{F}f$ where $\mathcal{F}$ is the Fourier transform and $f$ is an element of the associative $*$-algebra $\mathcal{M}$ modeling the $\rho$-Minkowski space{\footnote{Our convention for the Fourier transform is $\mathcal{F}f(p) = \int \frac{d^dx}{(2\pi)^d}\ e^{- i p x} f(x)$ and $f(x) = \int d^dp\ e^{i p x} \mathcal{F}f(p)$.}}, with star-product denoted as usual by the symbol $\star$. The convolution product and corresponding involution $^*$ are defined as \cite{williams2007crossed}, \cite{folland2016course} 
\begin{equation}
    (F\circ G)(s)
    = \int_{\mathcal{G}}\ d\mu(t) F(s t) G(t^{-1}),\  F^*(s)
    = {\overline{F}}(s^{-1})
    \label{convol-gene}
\end{equation}
for any $F,G\in \mathbb{C}(\mathcal{G})$, $s\in\mathcal{G}$, where $d\mu$ denotes the left-invariant Haar measure and the second relation holds true for {\it{unimodular}} group, which is the case here. ${\overline{F}}$ is the complex-conjugate of $F$. For detailed mathematical material, see 
e.g. \cite{williams2007crossed}, \cite{kappa-weyl}, \cite{kappa-weyl-bis}. The Weyl map $Q$ is defined by
\begin{equation}
    Q(f)
    = \pi(\mathcal{F}f)
    \label{weyl-operat}
\end{equation}
where, given a unitary representation of $\mathcal{G}$ $\pi_U:\mathcal{G}\to\mathcal{B}({\mathcal{H}})$, the induced $*$-representation of $\mathbb{C}(\mathcal{G})$ on $\mathcal{B}({\mathcal{H}})$, $\pi:\mathbb{C}(\mathcal{G})\to\mathcal{B}({\mathcal{H}})$ is defined by
\begin{equation}
   \pi(F)
   = \int d\mu(s) F(s) \pi_U(s)
   \label{inducedrep} 
\end{equation}
for any $F\in\mathbb{C}(\mathcal{G})$ and is bounded and non-degenerate{\footnote{Up to additional technical conditions which are not essential for the present talk.}} and satisfies $\pi(F\circ G)
    = \pi(F)\pi(G)$, $\pi(F)^\ddag
    = \pi(F^*)$. $\pi(F)^\ddag$ denotes the adjoint operator of $\pi(F)$.  But the Weyl map $Q:\mathcal{M}\to\mathcal{B}({\mathcal{H}})$ satisfies also $Q(f\star g)
    = Q(f) Q(g)$ and $(Q(f))^\ddag
    = Q(f^\dag)$. Putting all together, one obtains (see foornote 2)
\begin{align}
    f\star g
    = \mathcal{F}^{-1} (\mathcal{F}f \circ \mathcal{F}g), && 
    f^\dag
    = \mathcal{F}^{-1} (\mathcal{F}(f)^*).
    \label{star-prodetinvol}
\end{align}
In the present situation, the Lie algebra of coordinates is the Euclidean algebra $\mathfrak{e}(2)$ with corresponding Lie group being the Euclidean group involving the isometries of the 2-d Euclidean space. Restricting ourselves for convenience to the unimodular special Euclidean group, one thus retains
\begin{equation}
    \mathcal{G}
    := SE(2)
    = SO(2) \ltimes_\phi \mathbb{R}^2
    \label{g-rho}
\end{equation}
where $\phi:SO(2)\to\text{Aut}(\mathbb{R}^2)$ is simply the usual action of any matrix of $SO(2)$ on $\mathbb{R}^2$. Parametrising any element of $\mathcal{G}$ as
$(R(\rho p_0), \vec{p})$ where $R(\rho p_0)$ is a 2x2 rotation matrix with "angle" $\rho p_0$, one easily obtains the group law for $\mathcal{G}$ given by
\begin{align}
  (R(\rho{p_0}),\vec{p})\ (R(\rho q_0),\vec{q})
  &= \Big(R(\rho(p_0 + q_0)), \vec{p} + R(\rho{p_0})\vec{q} \Big), &
  \label{rho1} \\
  (R(\rho{p_0}), \vec{p})^{-1}
  &= \Big(R({-\rho p_0}), - R(-\rho{p_0})\vec{p} \Big), &
  \bbone = \bbone_2
  \label{rho2}.
\end{align}
This, combined with \eqref{convol-gene}, \eqref{star-prodetinvol} and using the fact that the Haar measure for $\mathcal{G}$ \eqref{g-rho} is $d\mu 
    = d^2p\ dp_0
    = d^3p$, yields after straightforward computation
 \begin{align}
    (f\star_\rho g)(x_0,\vec{x}) 
    &= \int \frac{dp_0}{2\pi}\ dy_0\ e^{-i p_0 y_0} f(x_0 + y_0, \vec{x}) g(x_0, R(-\rho p_0)\vec{x}),
    \label{star-rho} \\
    f^\dag(x_0,\vec{x})
    &= \int \frac{dp_0}{2\pi}\ dy_0\ e^{-i p_0 y_0} \overline{f}(x_0 + y_0, R(-\rho p_0) \vec{x} )
    \label{invol-rho}
\end{align}
for any $f, g \in\mathcal{M}$. A subscript $\rho$ is added to the symbol $\star$ from now on. Notice that it is understood that the associative $*$-algebra $\mathcal{M}$ modeling the $\rho$-Minkowski space must be enlarged as a multiplier algebra in order for the expressions to be well-defined. This is exemplified in [DS] for the case of $\kappa$-Minkowski.\\
From \eqref{star-rho}, \eqref{invol-rho} a simple calculation gives rise to the original coordinates Lie algebra
\begin{align}
    [x_0, x_1] = i \rho x_2, &&
    [x_0, x_2] = - i \rho x_1, &&
    [x_1, x_2] = 0.
    \label{coord-alg}
\end{align}
The algebra \eqref{coord-alg} is, by construction, 3-dimensional. A 4-dimensional algebra corresponding the a 4-dimensional $\rho$-Minkowski space is readily obtained by supplementing the generators $x_0,x_1,x_2$ with a {\it{central element}} $x_3$. The extension of the star-product to incorporate this extra coordinate is straightforward. One gets
\begin{align}
    (f\star_\rho g)(x_0,\vec{x},x_3)
    &=\int \frac{dp_0}{2\pi}\ dy_0\ e^{-i p_0 y_0} f(x_0 + y_0, \vec{x}, x_3) g(x_0, R(- \rho p_0) \vec{x}, x_3),
    \label{star-final} \\
    f^\dag(x_0, \vec{x}, x_3)
    &= \int \frac{dp_0}{2\pi}\ dy_0\ e^{- i p_0 y_0} \overline{f}(x_0 + y_0, R( - \rho p_0) \vec{x}, x_3),
    \label{invol-final}
\end{align}
for any $f,g\in\mathcal{M}$. From now on, a 4-dimensional situation is considered.\\

The Lebesgue integral $\int d^4x $ defines a trace w.r.t.\ the star-products \eqref{star-rho} and \eqref{star-final}. Indeed, one can easily realize that 
\begin{equation}
 \int d^4x\ (f\star_\rho g)(x)=\int d^4x\ (g\star_\rho f)(x),\label{cyclic}
\end{equation}
for any $f,g\in\mathcal{M}$, while the integral $\int d^4x$ defines a positive map $\int d^4x:\mathcal{M}^{+}\to\mathbb{R}^+$ where $\mathcal{M}^+$ denotes the set of positive elements of $\mathcal{M}_{\rho}$, since one verifies that
\begin{equation}
    \int d^4x\ (f\star_\rho f^\dag)(x)
    = \int d^4x\ f(x)\overline{f}(x)\ge0\label{norml2},
\end{equation}
stemming from
\begin{align}
    \int d^4x\ (f \star_\rho g^\dag)(x)
    = \int d^4x\ f(x) \overline{g}(x), &&
    \int d^4x\ f^\dag(x)
    = \int d^4\ \overline{f}(x).
    \label{formule1}
\end{align}
One observes that this trace is {\it{not}} twisted contrary to the natural trace arising in the description of the $\kappa$-Minkowski space \cite{kappa-weyl-bis}, \cite{algeb-kappa}. This later is sometimes called a "twisted trace" in the physics literature, which actually defines in the mathematical literature a KMS weight \cite{kustermans}. The present difference comes from the structure of each of the Lie groups related to the coordinates algebras. In the case of $\rho$-Minkowski, the corresponding group is unimodular while in the $\kappa$-Minkowski case, the group is the affine group which is not unimodular. Accordingly, a non trivial (non unit) modular function will show up in various place of the construction and in particular will modify the cyclicity relation \eqref{cyclic} as $ \int d^4x\ (f\star g)(x)=\int d^4x\ ((\sigma\triangleright g)\star f)(x)$ where the twist $\sigma$ is essentially defined by the modular function. Notice that the KMS weight is an ingredient in the Tomita-Takesaki modular theory \cite{takesaki} which plays an important role in the set-up of the thermal time hypothesis \cite{connes-rov}, \cite{connes-rov2}. For a very recent related study within $\kappa$-Minkowski space-time extending the analysis \cite{kappa-weyl-bis}, see \cite{kilian-therm}.\\

The construction of action functionals will use a Hilbert product defined by
\begin{equation}
    \langle f,g\rangle :=\int d^4x\ (f^\dag\star_\rho g)(x)=\int d^4x\ \overline{f}(x)g(x),\label{zehilbertprod}
\end{equation}
for any $f,g\in\mathcal{M}$ where the rightmost equality formally coincides with the usual $L^2$ product.

\section{$\rho$-deformed relativistic symmetries: The $\rho$-Poincar\'e algebra}\label{section3}

It appears that $\mathcal{M}$ is a left-module algebra over a deformed Poincar\'e (Hopf) algebra $\mathcal{P}_\rho$ \cite{gauge-rho} for the action of $\mathcal{P}_\rho$ on $\mathcal{M}$, $\varphi:\mathcal{P}_\rho\otimes\mathcal{M}\to\mathcal{M}$, $\varphi(t\otimes f):=t\triangleright f$ defined by
\begin{align}
    (P_\mu \triangleright f)(x) &= -i \partial_\mu f(x),\ (M_j \triangleright f)(x) = (\epsilon_{jk}^{l} x^{k}P_l \triangleright f )(x),\nonumber\\ (N_j \triangleright f) &= ( (x_0 P_j - x_j P_0) \triangleright f)(x)\label{action-p-m}.
    \end{align}
This amount in particular to verify that $t\triangleright(f\star g)=m(\Delta(t)(\triangleright\otimes\triangleright)(f\otimes g))$ and 
$\varphi \circ (\text{id}_H \otimes 1_\mathcal{M}) = 1_\mathcal{M} \circ \epsilon$, for any $t\in\mathcal{P}_\rho$, $f,g\in\mathcal{M}$, where $\Delta:\mathcal{P}_\rho\otimes\mathcal{P}_\rho$, $\epsilon:\mathcal{P}_\rho\to\mathbb{C}$ are respectively the coproduct and co-unit while $m:\mathcal{M}\otimes\mathcal{M}\to\mathcal{M} $ is the star-product equipping $\mathcal{M}$.
In \eqref{action-p-m}, the indices for the generators $(P_\mu, M_j,N_j)$ are defined as $\mu \in \{0,+,-,3 \}$ and $j \in \{ +,-,3 \}$
in which $M_j$ and $N_j$ denote respectively the rotations and boosts and the $P_\mu$'s denote the translations, with $M_\pm=M_1+\pm iM_2$, $N_\pm=N_1\pm iN_2$, $P_\pm=P_1\pm iP_2$.\\
First, it follows from \eqref{action-p-m} that the Lie algebra structure of the Poincar\'e symmetry is not deformed, namely one has
\begin{equation}
    \begin{split}
        &  [M_i,N_j] = i \epsilon_{ijk}N_k, \quad [M_i,M_j] = i \epsilon_{ijk} M_k, \quad [N_i,N_j] = - i \epsilon_{ijk} M_k, \quad [N_i,P_0 ] = iP_i \\
       &   [N_i, P_j] = i \delta_{ij}P_0, \quad [M_i,P_j] = i \epsilon_{ijk} P_k, \quad [P_\mu,P_\nu] = [M_j,P_0] = 0\label{rhom-Lie},
    \end{split}
\end{equation}
A tedious computation \cite{gauge-rho} leads to the conclusion that the deformed Hopf algebra structure of the Poincar\'e symmetry for which $\mathcal{M}$ behaves as a left-module algebra over it is given by
\begin{align}
\Delta(M_\pm) &= M_\pm \otimes \bbone + \mathcal{E}_\mp \otimes M_\pm,\ \Delta(N_\pm)= N_\pm \otimes \bbone + \mathcal{E}_\mp \otimes  N_\pm-\rho P_\pm\otimes M_3\nonumber\\
\Delta(M_3)&= M_3 \otimes \bbone + \bbone \otimes M_3,\ \Delta(N_3)=  N_3 \otimes \bbone + \bbone \otimes N_3-\rho P_3\otimes M_3\nonumber\\
\Delta(P_{0,3})&=P_{0,3}\otimes\bbone+\bbone\otimes P_{0,3},\  \Delta(P_\pm)=P_\pm\otimes\bbone+\mathcal{E}_\mp\otimes P_\pm,\label{coproduit}
\end{align}
\begin{equation}
    \Delta(\mathcal{E}_\pm)=\mathcal{E}_\pm\otimes\mathcal{E}_\pm,\ \mathcal{E_\pm}=e^{\pm i\rho P_0}\label{coproduit-E}
\end{equation}
\begin{equation}
\epsilon(P_\mu) = 0,\  \quad \epsilon(\mathcal{E}) = 1,\ 
           \epsilon(M_j)= \epsilon(N_j) = 0,\ j=\pm,3,\label{epsilon}
\end{equation}
\begin{align}
           S(P_0)&=-P_0,\ S(P_3)=-P_3,\ S(P_\pm) = - \mathcal{E}_\mp P_\pm,\  \quad S(\mathcal{E}) = \mathcal{E}^{-1} \nonumber\\
           S(M_j) &= - M_j, \quad S(N_j) = - N_j,\ j=\pm,3,\label{antipode}
\end{align}
where one has
\begin{equation}
    \mathcal{E}_\mu=\bbone,\bbone,e^{\pm i\rho P_0},\ \mu=0,3,\pm\label{cestlestwists}.
\end{equation}
which correspond to the twists which will appear in the twisted differential calculus constructed later on, while $S:\mathcal{P}_\rho\to\mathcal{P}_\rho$ \eqref{antipode} is the antipode verifying $m\circ(S\otimes\text{id})\circ\Delta=m\circ(\text{id}\otimes S)\circ\Delta=\eta \otimes\epsilon$, with unit $\eta: \mathbb{C} \to \mathcal{H}$.\\

Let $\mathcal{T}_\rho\subset\mathcal{P}_\rho$ be the Hopf subalgebra generated by the deformed translations. One can verify that $\mathcal{T}_\rho$ and $\mathcal{M}$ are dual as Hopf algebras for the dual pairing $\langle. ,.\rangle:\mathcal{T}_\rho\times \mathcal{M}\to \mathcal{M}$ given by
\begin{equation}
    \langle P_\mu,x_\mu\rangle=-i\delta_{\mu\nu}\label{zepairinghopf}.
\end{equation}
Indeed, a standard computation leads to the conclusion that
\begin{eqnarray}
    \langle\Delta(t),x\otimes y\rangle&=&\langle t,x\star y \rangle=\langle 
    t_{(1)},x\rangle\langle t_{(2)},x \rangle\label{poule1}\\
    \langle ht,x\rangle&=&\langle h\otimes t,\Delta_{\mathcal{M}}(x)\rangle=\langle h,x_{(1)}\rangle\langle t,x_{(2)} \label{poule2}\rangle
\end{eqnarray}
\begin{equation}
    \langle S(t),x\rangle=\langle t,S_{\mathcal{M}}(x)\rangle\label{ouf}
\end{equation}
holds true for any $t\in\mathcal{T}_\rho$, $x,y\in\mathcal{M}$, where the Sweedler notation has been used and $\Delta_{\mathcal{M}}$, $\epsilon_{\mathcal{M}}$ and $S_{\mathcal{M}}$ are respectively the coproduct, co-unit and antipode for the trivial Hopf algebra structure supported by $\mathcal{M}$ which is defined by
\begin{equation}
    \Delta_{\mathcal{M}}(x_\mu)=x_\mu\otimes\bbone+\bbone\otimes x_\mu,\ \ 
   \epsilon_{\mathcal{M}}(x_\mu)=0,\ \ S_{\mathcal{M}}(x_\mu)=-x_\mu\label{dual2}.
\end{equation}

Other $\rho$-deformations of the Poincar\'e algebra have been considered in the literature, see \cite{marija1}, \cite{marija2} and algebraically studied in \cite{fabiano2023bicrossproduct} where in particular two star products have been presented which however are different from the star product \eqref{star-final}. The possible relationship between the Hopf algebra $\mathcal{P}_\rho$ \cite{gauge-rho} and their counterparts considered in \cite{fabiano2023bicrossproduct} has not been investigated so far.\\

Finally, one can easily realize that any classical action functional of the form $\int d^dx\ \mathcal{L}$ where $\mathcal{L}$ is some smooth Lagrangian depending on (smooth) fields is invariant under the action of $\mathcal{P}_\rho$; namely, one has
\begin{equation}
    h\blacktriangleright\int d^4x\ \mathcal{L}:=\int d^4x\ h\triangleright \mathcal{L}=\epsilon(h)\int d^4x\ \mathcal{L},\label{decadixII}
\end{equation}
which is verified for any $h\in\mathcal{P}_\rho$, $\mathcal{L}\in\mathcal{M}$ where of course the action defined in \eqref{action-p-m} is still assumed and $\epsilon(h)$ is given by \eqref{epsilon}. \\

\section{One-loop exploration of $\phi^4$ scalar field theory 
on $\rho$-Minkowski}\label{section4}

It is instructive to study the one-loop behaviour of two types of (positive) action functionals in 4 dimensions, differing from each other by their quartic interaction, either orientable or non-orientable. Namely, the corresponding action functionals are respectively given by
\begin{align}
\begin{aligned}
    S(\phi, \overline{\phi})
    &= \langle \partial \phi, \partial \phi \rangle
    + m^2 \langle \phi, \phi \rangle
    + g \langle \phi^\dag \star_\rho \phi, \phi^\dag \star_\rho \phi \rangle \\
    &= \int d^4x\ (\partial_\mu \overline{\phi} \partial_\mu \phi + m^2 \phib \phi)
    + g \int d^4x\ \phi^\dag \star_\rho \phi \star_\rho \phi^\dag \star_\rho \phi, \label{action1}
\end{aligned}
\end{align}
and
\begin{align}
\begin{aligned}
    S(\phi, \overline{\phi})
    &= \langle \partial \phi, \partial \phi \rangle
    + m^2 \langle \phi, \phi \rangle
    + g \langle \phi \star_\rho \phi, \phi \star_\rho \phi \rangle \\
    &= \int d^4x\ (\partial_\mu \overline{\phi} \partial_\mu \phi + m^2 \phib \phi)
    + g \int d^4x\ (\phi^\dag \star_\rho \phi^\dag \star_\rho \phi \star_\rho \phi)(x),
\end{aligned}
    \label{action2}
\end{align}
where the fields $\phi$, $\overline{\phi}$ and the parameter $m$ have mass dimension 1 and $g$ is a dimensionless coupling constant.\\

\begin{figure}%[h]
    \begin{minipage}{.495\textwidth}
         \centering
    \begin{tikzpicture}[scale = 1.5]
        \draw[black] (-.8,-.1) rectangle (-.6,.1);
        \draw[black] ( .6,-.1) rectangle ( .8,.1);
        \draw[black] (-1.2, -.7) node[anchor= east]{$\phi (k_2)$} to (-.8, -.1);
        \draw[black] (-1.2,  .7) node[anchor= east]{$\phib(k_1)$} to (-.8,  .1);
        \draw[black] ( 1.2, -.7) node[anchor= west]{$\phib(k_3)$} to ( .8, -.1);
        \draw[black] ( 1.2,  .7) node[anchor= west]{$\phi (k_4)$} to ( .8,  .1);
        \draw[black,
            decoration={markings, mark=at position .5 with {\arrow{To}}},
            postaction={decorate}
            ] (-.6,  .1) to[out=45, in=135] (.6, .1);
        \node at (0,.6) {$k_5$};
        \draw[black,
            decoration={markings, mark=at position .5 with {\arrow{To}}},
            postaction={decorate}
            ] ( .6,  -.1) to[out=-135, in=-45] (-.6, -.1);
        \node at (0,-.6) {$k_6$};
        \draw (0,-1) node[anchor = north]{$V(1256)V(3465) = V(1256)V(5643)$} to (0,-1);
    \end{tikzpicture}
    \end{minipage}%
    \begin{minipage}{.495\textwidth}
         \centering
    \begin{tikzpicture}[scale = 1.5]
        \draw[black] (-.8,-.1) rectangle (-.6,.1);
        \draw[black] ( .6,-.1) rectangle ( .8,.1);
        \draw[black] (-1.2, -.7) node[anchor= east]{$\phi (k_4)$} to (-.8, -.1);
        \draw[black] (  .2,  .7) node[anchor= south west]{$\phib(k_3)$} to ( .6,  .1);
        \draw[black] ( -.2,  .7) node[anchor= south east]{$\phi (k_2)$} to (-.6,  .1);
        \draw[black] ( 1.2,  .7) node[anchor= west]{$\phib(k_1)$} to ( .8,  .1);
        \draw[black] (-.8,.1) to (-1.1, .6) to[out=135, in=135] (-1.1, -.5);
        \draw[black,
            decoration={markings, mark=at position .5 with {\arrow{To}}},
            postaction={decorate}
            ] (-1.05, -.55) to[out=-45, in=-45] (1.1, -.4) to (.8, -.1);
        \node at (.4,-.7) {$k_5$};
        \draw[black,
            decoration={markings, mark=at position .5 with {\arrow{To}}},
            postaction={decorate}
            ] ( .6,  -.1) to[out=-135, in=-45] (-.6, -.1);
        \node at (0,0) {$k_6$};
        \draw (0,-1) node[anchor = north]{$V(5462)V(3615)$} to (0,-1);
    \end{tikzpicture}
    \end{minipage}%
    
    \caption{Two diagrams contributing to the 4-point function -- Orientable interaction.}
    \label{fig:4_pt}
\end{figure}
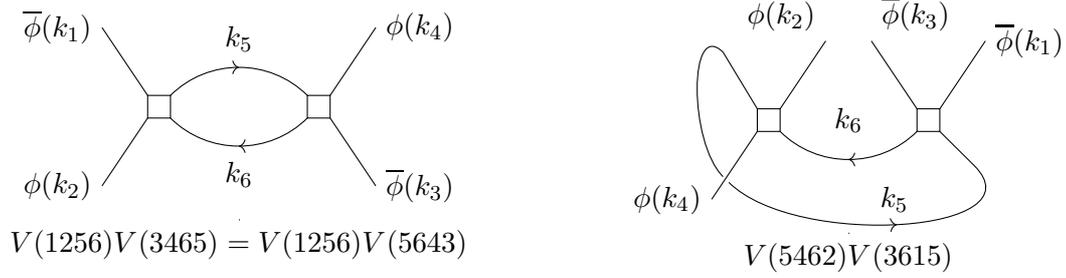

The computation of the one-loop contributions to the 2-point and 4-point functions in both cases is a simple routine computation. The results for the 2-point functions can be summarized as follows.\\
The one-loop contributions to the 2-point function for both scalar field theories exhibits a UV quadratic divergence as its commutative counterpart. As expected, the corresponding contributions are related to planar diagrams. In the orientable case \eqref{action1}, no IR singularities appears in the 2-point function which could generate UV/IR mixing while a IR singularity does appear in the case of non-orientable interaction \eqref{action2} thus signaling appearance of UV/IR mixing in this case. Note that a similar behaviour has been evidenced for 
orientable and non-orientable field theories in the $\kappa$-Minkowski case.\\
\begin{figure}%[h]
    \begin{minipage}{.249\textwidth}
         \centering
    \begin{tikzpicture}[scale = 1.2]
        \draw[black] (-.1,-.1) rectangle (.1,.1);
        \draw[black] (-.7, -.7) node[anchor= east]{$\phi (k_2)$} to (-.1, -.1);
        \draw[black] (-.7,  .7) node[anchor= east]{$\phib(k_1)$} to (-.1,  .1);
        \draw[-{To}, black] (.1, .1) to (.6, .6) to[out= 45, in= 90] (.9,0)
            node[anchor = east]{$k_3$};
        \draw[black] (.1,-.1) to (.6,-.6) to[out=-45, in=-90] (.9,0);
        \draw (0,-.8) node[anchor = north]{$V(1233)$} to (0,-.8);
    \end{tikzpicture}
    \end{minipage}%
    \begin{minipage}{.249\textwidth}
         \centering
    \begin{tikzpicture}[scale = 1.2]
        \draw[black] (-.1,-.1) rectangle (.1,.1);
        \draw[black] (.7,  .7) node[anchor= west]{$\phib(k_1)$} to (.1,  .1);
        \draw[black] (.7, -.7) node[anchor= west]{$\phi (k_2)$} to (.1, -.1);
        \draw[black] (-.1,-.1) to (-.6,-.6) to[out=-135, in=-90] (-.9,0);
        \draw[-{To}, black] (-.1, .1) to (-.6, .6) to[out= 135, in= 90] (-.9,0)
            node[anchor = west]{$k_3$};
        \draw (0,-.8) node[anchor = north]{$V(3312)$} to (0,-.8);
    \end{tikzpicture}
    \end{minipage}%
    \begin{minipage}{.249\textwidth}
         \centering
    \begin{tikzpicture}[scale = 1.2]
        \draw[black] (-.1,-.1) rectangle (.1,.1);
        \draw[black] (-.7, -.7) node[anchor= east]{$\phi (k_2)$} to (-.1, -.1);
        \draw[black] ( .7, -.7) node[anchor= west]{$\phib(k_1)$} to ( .1, -.1);
        \draw[-{To}, black] (-.1,  .1) to (-.6,  .6) to[out=135, in=180] (0, .9)
            node[anchor = north]{$k_3$};
        \draw[black] (.1,.1) to (.6,.6) to[out=45, in=0] (0, .9);
        \draw (0,-.8) node[anchor = north]{$V(3213)$} to (0,-.8);
    \end{tikzpicture}
    \end{minipage}%
    \begin{minipage}{.249\textwidth}
         \centering
    \begin{tikzpicture}[scale = 1.2]
        \draw[black] (-.1,-.1) rectangle (.1,.1);
        \draw[black] (-.7,.7) node[anchor= east]{$\phib(k_1)$} to (-.1,.1);
        \draw[black] ( .7,.7) node[anchor= west]{$\phi (k_2)$} to ( .1,.1);
        \draw[-{To}, black] (-.1, -.1) to (-.6, -.6) to[out=-135, in=180] (0, -.9)
            node[anchor = south]{$k_3$};
        \draw[black] (.1,-.1) to (.6,-.6) to[out=-45, in=0] (0, -.9);
        \draw (0,-.9) node[anchor = north]{$V(1332)$} to (0,-.9);
    \end{tikzpicture}
    \end{minipage}
    
    \caption{Diagrams contributing to the 2-point function -- Orientable interaction.}
    \label{fig:planar}
\end{figure}
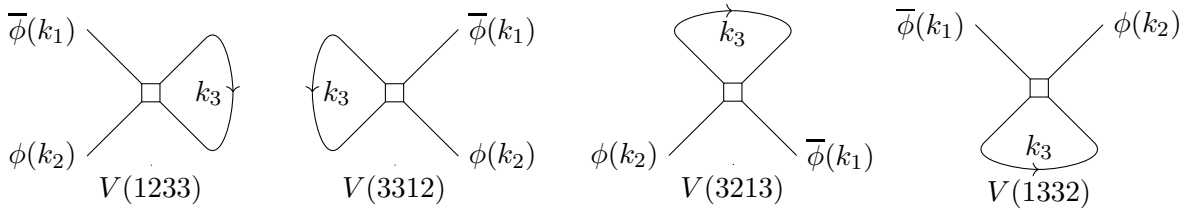
\begin{figure}%[h]
    \begin{minipage}{.24\textwidth}
         \centering
    \begin{tikzpicture}[scale = 1.2]
        \draw[black] (-.1,-.1) rectangle (.1,.1);
        \draw[black] (-.7, -.7) node[anchor= east]{$\phi (k_4)$} to (-.1, -.1);
        \draw[black] (-.7,  .7) node[anchor= east]{$\phib(k_3)$} to (-.1,  .1);
        \draw[-{To}, black] (.1, .1) to (.6, .6) to[out= 45, in= 90] (.9,0)
            node[anchor = east]{$k_1$};
        \draw[black] (.1,-.1) to (.6,-.6) to[out=-45, in=-90] (.9,0);
        \draw (0,-.8) node[anchor = north]{$V^{\mathrm{no}}(3411)$} to (0,-.8);
    \end{tikzpicture}
    \end{minipage}%
    \begin{minipage}{.24\textwidth}
         \centering
    \begin{tikzpicture}[scale = 1.2]
        \draw[black] (-.1,-.1) rectangle (.1,.1);
        \draw[black] (.7,  .7) node[anchor= west]{$\phib(k_3)$} to (.1,  .1);
        \draw[black] (.7, -.7) node[anchor= west]{$\phi (k_4)$} to (.1, -.1);
        \draw[black] (-.1,-.1) to (-.6,-.6) to[out=-135, in=-90] (-.9,0);
        \draw[-{To}, black] (-.1, .1) to (-.6, .6) to[out= 135, in= 90] (-.9,0)
            node[anchor = west]{$k_1$};;
        \draw (0,-.8) node[anchor = north]{$V^{\mathrm{no}}(1134)$} to (0,-.8);
    \end{tikzpicture}
    \end{minipage}%
    \begin{minipage}{.26\textwidth}
         \centering
    \begin{tikzpicture}[scale = 1.2]
        \draw[black] (-.1,-.1) rectangle (.1,.1);
        \draw[black] (-.7, -.7) node[anchor= east]{$\phi (k_4)$} to (-.1, -.1);
        \draw[black] ( .7,  .7) node[anchor= west]{$\phib(k_3)$} to ( .1,  .1);
        \draw[black,
            decoration={markings, mark=at position 0.75 with {\arrow{To}}},
            postaction={decorate}
        ] (-.1,  .1) to (-.6,  .6) to[out=135, in=135] (.45, .55);
        \node at (0,1) {$k_1$};
        \draw[black] ( .1, -.1) to ( .6, -.6) to[out=-45, in=-45] (.55, .45);
        \draw (0,-.8) node[anchor = north]{$V^{\mathrm{no}}(1431)$} to (0,-.8);
    \end{tikzpicture}
    \end{minipage}%
    \begin{minipage}{.26\textwidth}
         \centering
    \begin{tikzpicture}[scale = 1.2]
        \draw[black] (-.1,-.1) rectangle (.1,.1);
        \draw[black] (-.7,  .7) node[anchor= east]{$\phib(k_3)$} to (-.1,  .1);
        \draw[black] ( .7, -.7) node[anchor= west]{$\phi (k_4)$} to ( .1, -.1);
        \draw[black,
            decoration={markings, mark=at position 0.75 with {\arrow{To}}},
            postaction={decorate}
        ] ( .1,  .1) to ( .6,  .6) to[out=  45, in=  45] (.55, -.45);
        \node at (1,0) {$k_1$};
        \draw[black] (-.1, -.1) to (-.6, -.6) to[out=-135, in=-135] (.45, -.55);
        \draw (0,-.8) node[anchor = north]{$V^{\mathrm{no}}(3114)$} to (0,-.8);
    \end{tikzpicture}
    \end{minipage}
    
    \caption{Diagrams contribution to the 2-point function -- Non orientable interaction}
    \label{fig:no_vertex}
\end{figure}
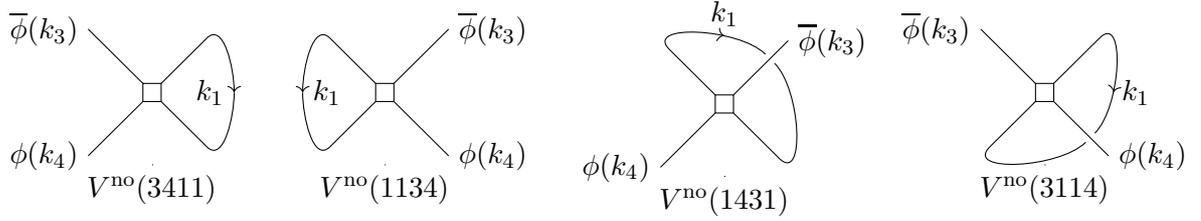
The one-loop 4-point function for the orientable scalar field theory \eqref{action1} has been studied in [nous] from the viewpoint of UV and IR behaviour. Typical diagrams among the twelve contributing diagrams are depicted on Figure \ref{fig:4_pt}.\\
One can show \cite{PW2019} that the one-loop contribution to the 4-point function has a logarithmic UV divergence as its commutative counterpart. Besides, there is an IR singularity showing up in some diagrams so that UV/IR mixing seems to be generated by 4-point contributions for orientable scalar field theories. This qualitatively agrees with the results of \cite{rho3}.\\

To conclude this section, UV/IR mixing shows up in both orientable and non-orientable scalar field theories on $\rho$-Minkowski space. This is to be compared with the absence of IR singularities in the 4-point function at one-loop within the orientable scalar field theory on $\kappa$-Minkowski space \cite{PW2019}, which by the way is also found to be UV finite. This model can thus be expected to be free of UV/IR mixing. Note however that the corresponding scalar propagator decays at large momenta "much faster" than the propagator for \eqref{action1}, \eqref{action2}, which explains the UV finitude of the one-loop 4-point function and the neutralization of dangerous IR singularities. For a recent reinvestigation of the UV/IR mixing, see \cite{kilian-uvir}.

\section{A gauge theory model on $\rho$-Minkowski}\label{section5}

It is convenient to use a twisted version of a noncommutative differential calculus based on the translation generators $P_\mu$. \\
Observe first that these later act as twisted derivations on $\mathcal{M}$, which is apparent from the expression for the coproduct 
$\Delta(P_\mu)$ \eqref{coproduit} or by explicitly computing the action of $P_\mu=-i\partial_\mu$ on $f\star g$ using \eqref{star-final}. Indeed, one arrives at 
\begin{eqnarray}
P_i(f\star g)&=&P_i(f)\star g+f\star P_i(g),\ i=0,3\nonumber\\
P_\pm(f\star g)&=&P_\pm(f)\star g+\mathcal{E}_\mp(f)\star P_\pm(g)\label{twist-leib},
\end{eqnarray}
where the twists $\mathcal{E}_\pm$ are defined in eqn. \eqref{cestlestwists}.
Consider now the following graded abelian Lie algebra of twisted derivations of $\mathcal{T}_\rho$ which is also a $\mathcal{Z}(\mathcal{M})$-bimodule{\footnote{One verifies $(z.P_\mu)(f)=z\star P_\mu(f)=P_\mu(f)\star z:=(P_\mu .z)(f),\ \mu=0,3,\pm$, for any $z\in\mathcal{Z}(\mathcal{M})$, the center of $\mathcal{M}$.}}
\begin{equation}
    \mathfrak{D}=\big\{P_\mu:\mathcal{M}_\rho\to\mathcal{M}_\rho,\ \mu=0,3,\pm,\ \{ P_0,P_3 \}_{\bbone}\oplus \{ P_+\}_{\mathcal{E}_+}\oplus \{ P_- \}_{\mathcal{E}_-}\big\}\label{chatte},
\end{equation}
where the grading is given by the twist degree defined by
\begin{equation}
    \tau(P_i)=0,\ i=0,3,\ \tau(P_\pm)=\pm1\label{twistdegree}.
\end{equation}
Accordingly, the linear structures in \eqref{chatte} are defined from {\it{homogeneous}} linear combinations of elements of \eqref{chatte}, i.e. all the elements of the linear combination have the same twist degree, which thus defines a grading which will extend to the differential calculus. This gives rise to the following relation in obvious notations
\begin{equation}
    \mathfrak{D}=\mathfrak{D}_0 \oplus \mathfrak{D}_+\oplus\mathfrak{D}_-\label{grading},
\end{equation}
where $\mathfrak{D}_i$, $i=0,\pm$ can be read off directly from \eqref{chatte}. Note that $\mathfrak{D}$ does not involve all the derivations of $\mathcal{M}$. It follows that the present differential calculi based on $\mathfrak{D}$ will be of restricted type as explained and formalized in \cite{dbv} and first used to construct gauge theories on various quantum spaces in \cite{cawa}, \cite{cawa2}.\\

In the present situation involving various type of twisted derivations and thus a grading, a noncommutative differential calculus is based on all the sets of $\mathcal{Z}(\mathcal{M})$-linear antisymmetric maps of degree $n$, says $\omega:\mathfrak{D}^n\to\mathcal{M}$. These sets are denoted by $\Omega^n_{(p,q,r)}(\mathcal{M})$ where the subscripts $p,q,r$ denote the respective integer twist degrees for $\{P_0,P_3\}$, $P_+$, $P_-$ such that $p+q+r=n$. A product of forms $\times$ must be defined as a map 
\begin{equation}
   \times:\Omega^n_{(p_1,q_1,r_1)}(\mathcal{M})\times\Omega^m_{(p_2,q_2,r_2)}(\mathcal{M})\to\Omega^{n+m}_{(p_1+p_2,q_1+q_2,r_1+r_2)}(\mathcal{M}).  
\end{equation}
As the ensuing analysis does not need to use explicitly the twist degree otherwise than to restrict the linear structures of $\mathfrak{D}$ to homogeneous structures with respect to this degree so that the above triplet of subscripts  will be omitted from now on and we set $\Omega^0(\mathcal{M})=\mathcal{M}$.\\

It can be shown \cite{gauge-rho} that the triplet $(\Omega^\bullet(\mathcal{M})=\bigoplus_{n=0}^4 \Omega^n(\mathcal{M}), \times, \text{d})$ defines a graded differential algebra of forms, where $\times$ and $\text{d}$ are respectively the product of forms and the differential. Any $n$-form $\omega\in\Omega^n(\mathcal{M})$ verifies
\begin{equation}
 \omega(P_1,P_2,...,P_n)\in\mathcal{M},\ \  \omega(P_1,P_2,...,P_n.z)=\omega(P_1,P_2,...,P_n)\star z  
\end{equation}
for any $z$ in $Z(\mathcal{M }_\rho)$ and $P_1,\dots,P_n\in\mathfrak{D}$, where the symbols $P_i$ stand for some derivations in $\mathfrak{D}$.\\    
For any $\omega\in\Omega^p(\mathcal{M})$, $\eta\in\Omega^q(\mathcal{M})$, the product $\times: \Omega^\bullet(\mathcal{M}) \to \Omega^\bullet(\mathcal{M})$ and differential $\text{d}$ are defined as   
\begin{equation}
\begin{aligned}
    (\omega \times \eta) & (P_1, \dots, P_{p+q}) \\
    &= \frac{1}{p!q!} \sum_{\sigma\in \kS_{p+q}} (-1)^{{\sign}(\sigma)}
    \omega(P_{\sigma(1)}, \dots, P_{\sigma(p)}) \star \eta(P_{\sigma(p+1)}, \dots, P_{\sigma(p+q)}),
    \label{eq:form_prod}
\end{aligned}
\end{equation}
\begin{align}
\begin{aligned}
    \text{d} \omega(P_1, \dots, P_{p+1}) 
    =& \sum_{j = 1}^{p+1} (-1)^{j+1} P_j \triangleright\big( \omega( P_1, \dots, \vee_j, \dots, P_{p+1}) \big),
    \label{eq:koszul}
\end{aligned}
\end{align}
where $\omega\times\eta\in\Omega^{p+q}(\mathcal{M})$, $\text{d}^2=0$, the symbols $P_i$, $i=1,2,\dots, p+q$ in \eqref{eq:form_prod}, \eqref{eq:koszul} denote generically some derivations $P_\mu,\ \mu=0,3,\pm\in\mathfrak{D}$, ${\sign}(\sigma)$ is the signature of the permutation $\sigma$, $\kS_{p+q}$ is the symmetric group of $p+q$ elements and the symbol $\vee_j$ denotes the omission of the element $j$. The differential satisfies the following twisted Leibnitz rule 
\begin{equation}
  \text{d}(\omega\times\eta)=\text{d}\omega\times\eta+(-1)^{\delta(\omega)} \omega\times_\mathcal{E}\text{d}\eta \label{leibnitz-form}
\end{equation}
for any $\omega,\eta\in\Omega^\bullet(\mathcal{M}_\rho)$, where the symbol $\delta(.)$ denotes the degree of form and the symbol $\times_\mathcal{E}$ indicates that a twist acts on the first factor. This twist is the one linked to the derivation acting on the 2nd factor in \eqref{leibnitz-form}. For more technical details, see [nous]. Observe that the above differential algebra is not graded commutative since one can verify that $\omega \times \eta \ne (-1)^{\delta(\omega)\, \delta(\eta)} \eta \times \omega$. Note that another example of suitable differential calculus has been presented in \cite{gauge-rho}.\\

In the present situation, a noncommutative version of a connection can be defined, {\it{\`a la Koszul}} \cite{dbv}, as a map $ \nabla:\mathfrak{D}_i\times\mathbb{E}\to\mathbb{E},\ i=0,\pm$, where $\mathbb{E}$ is a right hermitian module over the algebra $\mathcal{M}$ with a given action of the algebra
on the module, says $m\triangleleft f$, such that the following conditions holds
\begin{equation}
\nabla_{P_\mu+P^\prime_\mu}(m)=\nabla_{P_\mu}(m)+\nabla_{P^\prime_\mu}(m),\ \forall (P_\mu,P^\prime_\mu)\in\mathfrak{D}_i\times\mathfrak{D}_i, \ i=0,\pm\label{sigtaucon1}
\end{equation}
\begin{equation}
  \nabla_{z.P_\mu}(m)=\nabla_{P_\mu}(m)\star z,\ \forall P_\mu\in\mathfrak{D},\forall z\in Z(\mathcal{M }_\rho)\label{sigtaucon2},
\end{equation}
\begin{equation}
    \nabla_{P_\mu}(m\triangleleft f)=\nabla_{P_\mu}(m)\triangleleft f+{\beta}_{P_\mu}(m)\triangleleft P_\mu(f),\ \forall P_\mu\in\mathfrak{D}\label{sigtaucon3},
\end{equation}
in which \eqref{sigtaucon1} applies to linear combinations of derivations homogeneous in the twist degree \eqref{twistdegree} and ${\beta}_{P_\mu}:\mathbb{E}\to\mathbb{E}$ to be determined in a while.\\

The usual assumption in (most of) the physics literature relies in the choice
\begin{equation}
\mathbb{E}\simeq\mathcal{M}_\rho,\ \ m\triangleleft f=m\star f,\ \ h(m_1,m_2)=m_1^\dag\star m_2\label{unecopy}
\end{equation}
for any $m,m_1,m_2\in\mathbb{E}\simeq\mathcal{M}$ 
where $h:\mathbb{E}\times\mathbb{E}\to\mathcal{M}$ is the hermitian structure equipping the module which is one copy of the algebra. This assumption will be used in the sequel. Accordingly, the last condition \eqref{sigtaucon3} becomes
\begin{equation}
    \nabla_{P_\mu}(m\triangleleft f)=\nabla_{P_\mu}(m)\triangleleft f+\mathcal{E}_\mu(m)\triangleleft P_\mu(f),\ \forall P_\mu\in\mathfrak{D}\label{sigtaucon4}.
\end{equation}
where the map ${\beta}_{P_\mu}=\mathcal{E}_\mu$, $\mu=0,3,\pm$ is fixed by the requirement that the consistency relation $\nabla_{P_\mu}((m\star f)\star g)=\nabla_{P_\mu}(m\star (f\star g))$ holds true. \\

Upon setting $A_\mu=\nabla_{P_\mu}(\bbone),\ \nabla_\mu:=\nabla_{P_\mu}$, thus introducing the gauge potential, \eqref{sigtaucon4} gives rise to\\
\begin{equation}
   \nabla_{\mu}(f)=A_{\mu}\star f+ P_\mu(f)  \label{covar-deriv}
\end{equation}
for any $f\in\mathcal{M}$. Besides, compatibility between the connection and the hermitian structure can be expressed as twisted hermiticity conditions for $A_\pm$ and $A_0,A_3$ respectively given by
\begin{equation}
(h(\mathcal{E}_+\triangleright\nabla_+(m_1),m_2)
    +h(\mathcal{E}_+\triangleright m_1,\nabla_+(m_2)))+ (+\to -)=P_+h(m_1,m_2)+P_-h(m_1,m_2)\label{hermiticity-cond1},
\end{equation}
\begin{equation}
    h(\nabla_i(m_1),m_2)
    +h(m_1,\nabla_i(m_2)))=P_ih(m_1,m_2),\ i=0,3\label{hermiticity-cond2}
\end{equation}
for any $m_1,m_2\in\mathbb{E}$, which are verified provided
\begin{equation}
    A_\pm^\dag=\mathcal{E}_\pm\triangleright A_\mp,\ A_i^\dag=A_i,\ i=0,3\label{hermit-amu}.
\end{equation}

The curvature $\mathcal{F}(P_\mu,P_\nu):=\mathcal{F}_{\mu\nu}:\mathbb{E}\to\mathbb{E},\ \mu,\nu=0,3,\pm $ can be expressed as [nous]
\begin{equation}
    \mathcal{F}_{\mu \nu} : = \mathcal{E}_\nu \nabla_\mu \mathcal{E}_\nu^{-1} \nabla_\nu - \mathcal{E}_\mu \nabla_\nu \mathcal{E}_\mu^{-1} \nabla_\mu,\ \ \mu,\nu=0,3,\pm\label{morph-curv}.
\end{equation}
One easily verifies that $ \mathcal{F}_{\mu\nu}(m\star f)=\mathcal{F}_{\mu\nu}(m)\star f$ for any $m\in\mathbb{E}$, $f\in\mathcal{M}$, i.e. $ \mathcal{F}_{\mu\nu}$ defines a morphism of module. Then, upon setting $ \mathcal{F}_{\mu\nu}(\bbone)=F_{\mu\nu}  $, one obtains
\begin{equation}
F_{\mu \nu} = P_\mu A_\nu - P_\nu A_\mu + (\mathcal{E}_\nu\triangleright A_\mu) \star A_\nu - (\mathcal{E}_\mu\triangleright A_\nu) \star A_\mu,\ \ \mu,\nu=0,3\pm\label{fmunu},
\end{equation}
which can be viewed as a noncommutative analog of a "field strength".\\

The group of unitary gauge transformations is given by
\begin{equation}
    \mathcal{U}=\{g\in\mathbb{E}\simeq\mathcal{M}_\rho,\ \ g^\dag\star g= g\star g^\dag= 1  \}.\label{groupdejauge}
\end{equation}
To obtain \eqref{groupdejauge}, simply use the fact that the "noncommutative" unitary gauge transformations are defined as the automorphisms of $\mathbb{E}$ preserving the hermitian structure, i.e. $h(\phi(m_1),\phi(m_2))=h(m_1,m_2)$ for $\phi\in\text{Aut}(\mathbb{E})$ in obvious notations, combined with the last relation of \eqref{unecopy} and set $\phi(\bbone)=g$. The twisted gauge transformations for the connection are
\begin{equation}
    \nabla^g_\mu(.)=(\mathcal{E}_\mu\triangleright g^\dag)\star\nabla_\mu(g\star .)\label{gauge-connex},
\end{equation}
for any $g\in\mathcal{U}$, from which a standard computation leads to
\begin{equation}
A^g_\mu=(\mathcal{E}_\mu\triangleright g^\dag)\star A_\mu\star g+(\mathcal{E}_\mu\triangleright g^\dag)\star P_\mu g\label{gaugetrans-amiou},
\end{equation}
\begin{equation}
    F^g_{\mu\nu}=(\mathcal{E}_\mu\mathcal{E}_\nu\triangleright g^\dag)\star F_{\mu\nu} \star g 
    \label{transfmunu},
\end{equation}
where there is no summation over indices $\mu,\nu$ in the RHS of \eqref{transfmunu} and the twists $\mathcal{E}_\mu$ are still given by \eqref{cestlestwists}.\\

A gauge invariant action functional whose formal commutative limit $\rho\to0$ is equal to the standard action functional of 4-d QED can be easily derived \cite{gauge-rho} from the combination of \eqref{zehilbertprod} and \eqref{fmunu}. It is simply given by
\begin{equation}
    S_\rho:=\frac{1}{4G^2}\langle F_{\mu\nu}, F_{\mu\nu}\rangle =\frac{1}{4G^2}\int d^4x\  F_{\mu\nu}^\dag\star F_{\mu\nu}=\frac{1}{4G^2}\int d^4x\ \overline{F_{\mu\nu}}(x) F_{\mu\nu}(x)\label{action-class},
\end{equation}
where $G$ is a dimensionless coupling constant and summation over $\mu, \nu$ is understood. By further assuming that the components of the gauge potential $A_0, A_1, A_2,A_3$ are real valued, \eqref{action-class} obviously reduces to the standard action for QED when $\rho\to0$. Besides, upon using the cyclicity of the Lebesgue integral (i.e. the natural trace here) \eqref{cyclic} together with the "unitary relation" in \eqref{groupdejauge} and the relations
\begin{eqnarray}
    ((\mathcal{E}_\mu\mathcal{E}_\nu)\triangleright g)^\dag&=&(\mathcal{E}_\mu
    \mathcal{E}_\nu)\triangleright g,\ \mu,\nu=0,3,\pm,\nonumber\\
   \mathcal{E}_\mu\triangleright(f\star g)&=&(\mathcal{E}_\mu\triangleright f)\star(\mathcal{E}_\mu\triangleright g)  
\end{eqnarray}
for any $g\in\mathcal{U}$, one easily verifies that \eqref{action-class} is invariant under the gauge tranformations \eqref{groupdejauge}.\\
Finally, as a direct consequence of \eqref{decadixII}, the action functional \eqref{action-class} is invariant under the $\rho$-Poincar\'e symmetry. Notice that the above construction can be straightforwardly adapted to a 3-dimensional situation.\\
One can verify that the kinetic operator of \eqref{action-class} coincides with the one of usual electrodynamics, i.e. $S_{\rho,kin}\sim\int d^4x\ A_\mu( \eta_{\mu\nu}\partial^2-\partial_\mu\partial_\nu)A_\nu$ (where the sum is over $\mu,\nu=0,1,2,3$), which comes from \eqref{zehilbertprod} and the chosen set of twisted derivations \eqref{chatte}.

\section{Discussion and outlook}\label{section6}

A star-product realization of the $\rho$-Minkowski space-time can be 
obtained from a combination of the Weyl quantization map and the defining properties of the convolution algebra of the special Euclidean group which is the Lie group related to the coordinates algebra, in a way similar to what has been done for $\kappa$-Minkowski \cite{kappa-weyl}, \cite{kappa-weyl-bis}. Both quantum space-times are of course acted on by a deformation of the Poincar\'e symmetry modeled by a deformed Hopf algebra whose subalgebra of "deformed translations", acting as twisted derivations, is a natural candidate to enter the construction in each case of a (noncommutative) differential calculus \cite{dbv} based on these twisted derivations.\\

Gauge theory models either for $\rho$-Minkowski \cite{gauge-rho} or for $\kappa$-Minkowski \cite{MW2020a} have been obtained using the above framework. The main differences between both actions functionals at the classical level are essentially of algebraic origin, stemming from the twists affecting the translations/twisted derivations or from the unimodular or non-unimodular nature of the group linked to the coordinate algebras. The nature of the twisted derivations impacts the differential calculus and by the way fixes the expression for the kinetic operator of the action functional. While a unimodular Lie group for coordinates as in the $\rho$-Minkowski case \cite{gauge-rho} implies the existence of a natural trace as the Haar measure of the group, which is simply the Lebesgue integral here, this simple situation is modified whenever the group is non-unimodular as the modular function comes into play \cite{williams2007crossed}, \cite{folland2016course}, as in the $\kappa$-Minkowski case \cite{MW2020a}. The trace must be trade for a "twisted trace", known in the mathematical literature as a KMS weight \cite{kustermans}, which is still the Lebesgue integral in the $\kappa$-Minkowski case, whose cyclicity w.r.t. the star-product is altered by a twist in close relationship with the Tomita-Takesaki modular group \cite{takesaki}. Both traces are natural ingredients in the construction of a gauge invariant action functional in each of these quantum space-times, with however a strong constraint on the dimension of the space-time in the $\kappa$-Minkowski which is fixed by gauge invariance to the unique value{\footnote{Recall that this comes from the requirement of gauge invariance of the action functional combined with the "modified cyclicity" of the twisted trace by a twist depending on the space-time $d$, resulting in the unique dimension $d=5$.}}$d=5$ , while in the $\rho$-Minkowski case, one has $d\ge3$, stemming simply from the nature of the algebra \eqref{coord-alg}. Finally, notice that both actions functionals are invariant under the deformed-Poincar\'e symmetries, which is a mere consequence of the construction, stemming fromthe use of the Lebesgue integral corresponding in each case to the Haar measure (right-invariant measure in the non-unimodular case).\\

As far as one-loop corrections are concerned for the action functional \eqref{action-class}, one has first to compute the 1-point tadpole function whose non-vanishing may signal a vacuum instability under quantum fluctuations, which would imply, at least, a radiative breaking of the gauge invariance (as well as Lorentz symmetry breaking). The computation is done by first adding to $S$ \eqref{action-class} a gauge-fixing action resulting in
\begin{align}
S&=S_\rho+s\int d^4x\ ({\overline{C}}^\dag\star({\overline{P_\mu}}\triangleright(A_\mu)-\lambda)
+c.c.,\ \ (s^2=0)\label{actionfixed}
\end{align}
where a noncommutative analog of the Lorentz gauge has been choosen. Namely one has, for $\lambda(x)=0$, the following relation ${\overline{P_\mu}}\triangleright(A_\mu)=P_0A_0+P_3A_3+P_+A_-+P_-A_+$, ad $s$ is the nilpotent Slavnov operation defining the (twisted) BRST symmetry defined as in \cite{brst-kappa}, whose structure equations in the present situation are
\begin{align}
sA_\mu&=P_\mu C+A_\mu\star C-(\mathcal{E}_\mu\triangleright C)\star A_\mu\nonumber \\
s\overline{C}^\dag&=b^\dag,\ \ sb^\dag=0
\end{align}
supplemented by $s\lambda=0$. A tedious computation indicates unfortunately that the 1-point function is proportional to $\lambda$ and is thus non-vanishing, as it has been shown for the $\kappa$-Minkowski case \cite{tadpole}. Note that the appearance of a non-vanishing tadpole is a frequent phenomenon in gauge theories on quantum spaces, see e.g. \cite{R2}, \cite{R4} respectively corresponding to a gauge matrix model on the 2-d Moyal space and a Yang-mills type model as the one for $\rho$- or $\kappa$-Minkowski built or a deformation of the 3-d Euclidean space.\\
It is not known if one could accommodate to a non-vanishing tadpole in these noncommutative gauge theories for interesting physical purpose related to some effective regime (to be defined) of Quantum Gravity. Besides, it is not known if this particular feature could be avoided within a class of Minkowski deformations so far unexplored from the viewpoint of field theory. Note that computing perturbative quantum corrections would no longer make sense if some of these noncommutative gauge theories was to be considered as some effective theories {\t{on}} a quantum space, which would definitely need to characterize its relationship to a more fundamental Quantum Gravity theory as a theory {\it{of}} space and the energy scale at which the effective theory operates. These points obviously raise challenging albeit fascinating issues.

\vfill\eject

{\bf{Acknowledgment:}} I thank the organizers of the Corfu Summer Institute 2024 and the Workshop on Noncommutative and Generalized Geometry in String theory, Gauge theory and Related Physical Models for their invitation and the Action 21109 CaLISTA ``Cartan geometry, Lie, Integrable Systems, quantum group Theories for Applications'', from the European Cooperation in Science and Technology. Discussions with P. Bieliavsky during the workshop are gratefully acknowledged. I thank V. Maris for constructive exchanges and comments.

\end{document}